# CROSS-SECTION AND SELECTION RULES IN SURFACE ENHANCED HYPER RAMAN SCATTERING

**A.M. Polubotko**

A.F. Ioffe Physico-Technical Institute Russian Academy of Sciences, Politechnicheskaya 26, 194021 Saint Petersburg, Russia, E-mail: alex.marina@mail.ioffe.ru

Tel: (812) 292-71-73, Fax: (812) 297-10-17

## Abstract

The expression for the SEHR cross-section of symmetrical molecules within the framework of the dipole-quadrupole theory is obtained. It is formed by contributions, which depend on various dipole and quadrupole moments. Estimation of the enhancement coefficients for the quadrupole enhancement mechanism in some limited case can achieve the value $10^{30}$. It is demonstrated that the contributions follow some selection rules. Qualitative classification of the contributions after the enhancement degree is given.

# 1. Introduction

Investigation of the reasons of the enhancement of optical processes on rough metal surfaces is of great importance due to large perspectives of using of these phenomena in chemistry, biology, physics and medicine. There are several points of view on the nature of these processes. Main opinion, which is widespread in literature is that the reason of the enhancement are surface plasmons and a charge transfer or chemical mechanism. We do not agree with this point of view and corresponding criticism is contained in[1-3]. Our opinion one can find in a recently published book[1]. At present it is well established that one of the main reasons of the enhancement is a large increase of the electromagnetic field and its derivatives near sharp points of rough surfaces. The second essential reason is a quantum mechanical feature of the quadrupole interaction of light with molecules[1-7]. This mechanism was well investigated for the SERS phenomenon, however it is of great importance to corroborate it on another processes. Here, in this first paper we obtain the general expression for the SEHR cross-section of symmetrical molecules. One can see, that it is determined by several contributions, which describe the scattering via various combinations of the dipole and quadrupole moments. The symmetry analysis allows us to establish selection rules for the contributions and then classify them after enhancement degree on the base of the estimations of the enhancement of dipole and quadrupole light-molecule interactions. It appears that the SEHR spectra of molecules with $C_{nh}$, $D$ and higher symmetry groups contain strong lines, caused by totally symmetric vibrations, transforming after the unit irreducible representations of corresponding symmetry groups. The second paper is devoted to the analysis of the enhancement of the contributions and interpretation of the SEHR spectra of the set of symmetrical molecules. It is demonstrated that this analysis strongly confirms our point of view.

First observation of SEHRS (from $SO_3^{2-}$ adsorbed on $Ag$ powder) was made in[8]. Later the authors in[9-13] and some others observed the SEHR spectra of various molecules and investigated various characteristics of this process. Their main topic was the study of various experimental details of SEHRS. From our point of view the most valuable information which allows to establish the mechanism of SEHRS is existence of allowed and forbidden lines in the SEHR spectra of symmetrical molecules. It allows to make an unequivocal conclusion about significant role of the quadrupole interaction in SEHRS. Therefore it appears that the most valuable for us papers are[14-16], which contain investigation of symmetrical properties of the SEHR spectra of pyrazine and phenazine, that allow to confirm our ideas about SEHRS mechanism. In addition analysis of the SEHR spectra of trans-1, 2-bis (4-pyridyle) ethylene, pyridine and crystal violet demonstrates the possibility of explanation of them on the base of the dipole-quadrupole theory too.

## 2. The expression for the SEHR cross-section of symmetrical molecules

The expression for the SEHR cross-section of symmetrical molecules can be obtained by the same manner as the cross-section for SERS[1]. Since SEHRS is a three quantum process, its cross-section can be obtained using the time dependent quantum mechanical perturbation theory. It is expressed via the

$$\frac{d}{dt}\left|w^{(3)}_{(n,\overline{V}\pm 1),(n,\overline{V})}(t)\right|^2, \tag{1}$$

where $w^{(3)}_{(n,\overline{V}\pm 1),(n,\overline{V})}(t)$ is the third term in expansion of the perturbation coefficient between the states with $(n,\overline{V}\pm 1)$ and $(n,\overline{V})$ quantum numbers and refers to the Stokes and AntiStokes processes respectively. Here $n$ is a quantum number of the ground state of the electron system, while $\overline{V}$ is the totality of all vibrational quantum numbers of molecular vibrations $(V_1, V_2...V_s...)$. The expression $(\overline{V}\pm 1)$ means the change of one of vibrational quantum number on one unit.

$$w^{(3)}_{(n,\overline{V}\pm 1),(n,\overline{V})}(t) = \left(-\frac{i}{\hbar}\right)^3 \sum_{\substack{m \\ m\neq n}} \left[ \int_0^t \langle n,\overline{V}\pm 1|\hat{H}^{inc}_{e-r} + \hat{H}^{scat}_{e-r}|k,\overline{V}\pm 1\rangle dt_1 \times \right.$$

$$\times \int_0^{t_1} \langle k,\overline{V}\pm 1|\hat{H}^{inc}_{e-r} + \hat{H}^{scat}_{e-r}|m,\overline{V}\pm 1\rangle dt_2 \times \int_0^{t_2} \langle m,\overline{V}\pm 1|\hat{H}^{inc}_{e-r} + \hat{H}^{scat}_{e-r}|n,\overline{V}\rangle dt_3 +$$

$$+ \int_0^t \langle n,\overline{V}\pm 1|\hat{H}^{inc}_{e-r} + \hat{H}^{scat}_{e-r}|k,\overline{V}\pm 1\rangle dt_1 \times$$

$$\times \int_0^{t_1} \langle k,\overline{V}\pm 1|\hat{H}^{inc}_{e-r} + \hat{H}^{scat}_{e-r}|m,\overline{V}\rangle dt_2 \times \int_0^{t_2} \langle m,\overline{V}|\hat{H}^{inc}_{e-r} + \hat{H}^{scat}_{e-r}|n,\overline{V}\rangle dt_3 +$$

$$+ \int_0^t \langle n,\overline{V}\pm 1|\hat{H}^{inc}_{e-r} + \hat{H}^{scat}_{e-r}|k,\overline{V}\rangle dt_1 \times$$

$$\left. \times \int_0^{t_1} \langle k,\overline{V}|\hat{H}^{inc}_{e-r} + \hat{H}^{scat}_{e-r}|m,\overline{V}\rangle dt_2 \times \int_0^{t_2} \langle m,\overline{V}|\hat{H}^{inc}_{e-r} + \hat{H}^{scat}_{e-r}|n,\overline{V}\rangle dt_3 \right]. \quad (2)$$

Where $\hat{H}^{inc}_{e-r}$ and $\hat{H}^{scat}_{e-r}$ are light-molecule interaction Hamiltonians of the incident and scattered fields, which have the form

$$\hat{H}^{inc}_{e-r} = |\overline{E}_{inc}| \frac{(\overline{e}^* \overline{f}_e^*)_{inc} e^{i\omega_{inc} t} + (\overline{e}\,\overline{f}_e)_{inc} e^{-i\omega_{inc} t}}{2}, \quad (3)$$

$$\hat{H}^{scat}_{e-r} = |\overline{E}_{scat}| \frac{(\overline{e}^* \overline{f}_e^*)_{scat} e^{i\omega_{scat} t} + (\overline{e}\,\overline{f}_e)_{scat} e^{-i\omega_{scat} t}}{2}. \quad (4)$$

Here $\overline{E}_{inc}$ and $\overline{E}_{scat}$ are vectors of the incident and scattered electric fields, $\omega_{inc}$ and $\omega_{scat}$ are corresponding frequencies, $\overline{e}$ -are polarization vectors of corresponding incident or scattered fields,

$$f_{e\alpha} = d_{e\alpha} + \frac{1}{2E_\alpha} \sum_\beta \frac{\partial E_\alpha}{\partial x_\beta} Q_{e\alpha\beta} \quad (5)$$

is an $\alpha$ component of the generalized vector of interaction of light with molecule,

$$d_{e\alpha} = \sum_i e x_{i\alpha} \qquad (6)$$

$$Q_{e\alpha\beta} = \sum_i e x_{i\alpha} x_{i\beta} \qquad (7)$$

are the $\alpha$ component of the dipole moment vector and the $\alpha\beta$ component of the quadrupole moments tensor of interaction of light with electrons of the molecule. Here under $x_{i\alpha}$ and $x_{i\beta}$ we mean coordinates $x, y, z$ of $i$ electron. Further we shall omit the index $e$ in the moments designations. The expressions for the wavefunctions have the form[1]

$$\Psi_{n\bar{V}} = \left[ \Psi_n^{(0)} + \sum_{\substack{l \\ l \neq n}} \frac{\sum_{s,p} R_{nl(s,p)} \sqrt{\frac{\omega_{(s,p)}}{\hbar}} \xi_{(s,p)} \Psi_l^{(0)}}{(E_n^{(0)} - E_l^{(0)})} \right] \alpha_{\bar{V}} \exp-(iE_{n\bar{V}} t)/\hbar, \qquad (8)$$

$$\Psi_{m\bar{V}} = \left[ \Psi_m^{(0)} + \sum_{\substack{k \\ k \neq m}} \frac{\sum_{s,p} R_{mk(s,p)} \sqrt{\frac{\omega_{(s,p)}}{\hbar}} \xi_{(s,p)} \Psi_k^{(0)}}{(E_m^{(0)} - E_k^{(0)})} \right] \alpha_{\bar{V}} \exp-(iE_{m\bar{V}} t)/\hbar \qquad (9)$$

Here the expression (8) is written for the ground state, while (9) for excited states. $\Psi_n^{(0)}$, $\Psi_m^{(0)}$, $E_n^{(0)}$ and $E_l^{(0)}$ are the wavefunctions and the energies of the molecule in the ground and excited states respectively, which are obtained in supposition of the motionless nuclei. $R_{n,l,(s,p)}$ are the coefficients, which describe excitation of the state $l$ from the ground state $n$ by the $(s, p)$ vibrational mode. $s$ numerates the group of degenerate vibrational states, while $p$, the states inside the group. More precise description of $R_{n,l,(s,p)}$ one can find in[1]. $\omega_{(s,p)}$ and $\xi_{(s,p)}$ are the frequencies and normal coordinates of the $(s, p)$ vibrational mode.

$$E_{m,\bar{V}} = E_m^{(0)} + \sum_{s,p} \hbar \omega_{(s,p)} (V_{(s,p)} + 1/2) \qquad (10)$$

is the full energy of the electron shell and vibrations in the $m$ electron state.

$$\alpha_{\bar{V}} = \prod_{(s,p)} N_{(s,p)} H_{V_{(s,p)}}\left(\sqrt{\frac{\omega_{(s,p)}}{\hbar}}\xi_{(s,p)}\right) \exp\left(-\frac{\omega_{(s,p)}\xi^2_{(s,p)}}{2\hbar}\right) \quad (11)$$

is the wavefunction of the nuclei vibrations. Here $N_{(s,p)}$ are normalization coefficients, $H_{V_{(s,p)}}$ - Hermitian polynomials. All other designations are conventional. Here and further we assume, that the vibrational spectrum and hence the frequencies and normal coordinates are the same for the ground and excited electronic states of the molecule. Using the above expressions one can obtain the SEHR cross-section for the $(s, p)$ vibrational mode in the form

$$d\sigma_{SEHRS_{(s,p)}}\binom{St}{AnSt} = \frac{\omega_{inc}\omega^3_{scat}}{64\pi^2\hbar^4\varepsilon_0^2 c^4}\left|\bar{E}_{inc}\right|^2_{vol}\frac{\left|\bar{E}_{inc}\right|^2_{surf}\left|\bar{E}_{inc}\right|^2_{surf}\left|\bar{E}_{scat}\right|^2_{surf}}{\left|\bar{E}_{inc}\right|^2_{vol}\left|\bar{E}_{inc}\right|^2_{vol}\left|\bar{E}_{scat}\right|^2_{vol}} \times$$

$$\times \left(\frac{\frac{V_{(s,p)}+1}{2}}{\frac{V_{(s,p)}}{2}}\right) \times \left|C_{V_{(s,p)}}\left[(\bar{e}\bar{f})_{inc},(\bar{e}\bar{f})_{inc},(\bar{e}^*\bar{f}^*)_{scat}\right]\binom{St}{anSt}\right|^2 dO.$$

(12)

The full SERS cross-section is

$$d\sigma_{SEHRS_s} = \sum_p d\sigma_{SEHRS_{(s,p)}}. \quad (12a)$$

Here in (12) the signs $surf$ and $vol$ mean, that the corresponding fields refer to the ones at the surface and in the volume. $(\bar{E}_{inc})_{surf}$ is the surface field generated by the plane wave $(\bar{E}_{inc})_{vol}$, while $(\bar{E}_{scat})_{surf}$ is the surface field, generated by the field $(\bar{E}_{scat})_{vol}$, which is an incident field from the direction for which the direction of the wave reflected from the surface coincides with the direction of the scattering.

$$C_{V_{(s,p)}}[f_1, f_2, f_3]\binom{St}{AnSt} = \sum_{m,r,l \neq n} \frac{\langle n|f_3|m\rangle\langle m|f_2|r\rangle\langle r|f_1|l\rangle R_{n,l,(s,p)}}{(E_n^{(0)} - E_l^{(0)})(\omega_{m,n} - 2\omega_{inc})(\omega_{r,n} \pm \omega_{(s,p)} - \omega_{inc})} +$$

$$+ \sum_{m,r,l \neq n} \frac{\langle n|f_2|m\rangle\langle m|f_3|r\rangle\langle r|f_1|l\rangle R_{n,l,(s,p)}}{(E_n^{(0)} - E_l^{(0)})(\omega_{m,n} + \omega_{scat} - \omega_{inc})(\omega_{r,n} \pm \omega_{(s,p)} - \omega_{inc})} +$$

$$+ \sum_{m,r,l \neq n} \frac{\langle n|f_2|m\rangle\langle m|f_1|r\rangle\langle r|f_3|l\rangle R_{n,l,(s,p)}}{(E_n^{(0)} - E_l^{(0)})(\omega_{m,n} + \omega_{scat} - \omega_{inc})(\omega_{r,n} \pm \omega_{(s,p)} + \omega_{scat})} +$$

$$+ \sum_{m,r,l \neq n} \frac{\langle n|f_3|m\rangle\langle m|f_2|r\rangle R^*_{r,l,(s,p)}\langle l|f_1|n\rangle}{(E_r^{(0)} - E_l^{(0)})(\omega_{m,n} - 2\omega_{inc})(\omega_{r,n} \mp \omega_{(s,p)} - \omega_{inc})} +$$

$$+ \sum_{m,r,l \neq n} \frac{\langle n|f_2|m\rangle\langle m|f_3|r\rangle R^*_{r,l,(s,p)}\langle l|f_1|n\rangle}{(E_r^{(0)} - E_{(l}^{(0)})(\omega_{m,n} + \omega_{scat} - \omega_{inc})(\omega_{r,n} \mp \omega_{(s,p)} - \omega_{inc})} +$$

$$+ \sum_{m,r,l \neq n} \frac{\langle n|f_2|m\rangle\langle m|f_1|r\rangle R^*_{r,l,(s,p)}\langle l|f_3|n\rangle}{(E_r^{(0)} - E_l^{(0)})(\omega_{m,n} + \omega_{scat} - \omega_{inc})(\omega_{r,n} \mp \omega_{(s,p)} + \omega_{scat})} +$$

$$+ \sum_{m,r,l \neq n} \frac{\langle n|f_3|m\rangle\langle m|f_2|l\rangle R_{r,l,(s,p)}\langle r|f_1|n\rangle}{(E_r^{(0)} - E_l^{(0)})(\omega_{m,n} \pm \omega_{(s,p)} - 2\omega_{inc})(\omega_{r,n} - \omega_{inc})} +$$

$$+ \sum_{m,r,l \neq n} \frac{\langle n|f_2|m\rangle\langle m|f_3|l\rangle R_{r,l,(s,p)}\langle r|f_1|n\rangle}{(E_r^{(0)} - E_l^{(0)})(\omega_{m,n} + \omega_{scat} \pm \omega_{(s,p)} - \omega_{inc})(\omega_{r,n} - \omega_{inc})} +$$

$$+ \sum_{m,r,l \neq n} \frac{\langle n|f_2|m\rangle\langle m|f_1|l\rangle R_{r,l,(s,p)}\langle r|f_3|n\rangle}{(E_r^{(0)} - E_l^{(0)})(\omega_{m,n} + \omega_{scat} \pm \omega_{(s,p)} - \omega_{inc})(\omega_{r,n} + \omega_{scat})} +$$

$$+ \sum_{m,r,l \neq n} \frac{\langle n|f_3|m\rangle R^*_{m,l,(s,p)}\langle l|f_2|r\rangle\langle r|f_1|n\rangle}{(E_m^{(0)} - E_l^{(0)})(\omega_{m,n} \mp \omega_{(s,p)} - 2\omega_{inc})(\omega_{r,n} - \omega_{inc})} +$$

$$+ \sum_{m,r,l \neq n} \frac{\langle n|f_2|m\rangle R^*_{m,l,(s,p)}\langle l|f_3|r\rangle\langle r|f_1|n\rangle}{(E_m^0 - E_l^0)(\omega_{m,n} + \omega_{scat} \mp \omega_{(s,p)} - \omega_{inc})(\omega_{r,n} - \omega_{inc})} +$$

$$+ \sum_{m,r,l \neq n} \frac{\langle n|f_2|m\rangle R^*_{m,l,(s,p)}\langle l|f_1|r\rangle\langle r|f_3|n\rangle}{(E_m^{(0)} - E_l^{(0)})(\omega_{m,n} + \omega_{scat} \mp \omega_{(s,p)} - \omega_{inc})(\omega_{r,n} + \omega_{scat})} +$$

$$+ \sum_{m,r,l \neq n} \frac{\langle n|f_3|l\rangle R_{m,l,(s,p)} \langle m|f_2|r\rangle \langle r|f_1|n\rangle}{(E_m^{(0)} - E_l^{(0)})(\omega_{m,n} - 2\omega_{inc})(\omega_{r,n} - \omega_{inc})} +$$

$$+ \sum_{m,r,l \neq n} \frac{\langle n|f_2|l\rangle R_{m,l,(s,p)} \langle m|f_3|r\rangle \langle r|f_1|n\rangle}{(E_m^{(0)} - E_l^{(0)})(\omega_{m,n} + \omega_{scat} - \omega_{inc})(\omega_{r,n} - \omega_{inc})} +$$

$$+ \sum_{m,r,l \neq n} \frac{\langle n|f_2|l\rangle R_{m,l,(s,p)} \langle m|f_1|r\rangle \langle r|f_1|n\rangle}{(E_m^{(0)} - E_l^{(0)})(\omega_{m,n} + \omega_{scat} - \omega_{inc})(\omega_{r,n} + \omega_{scat})} +$$

$$+ \sum_{m,r,l \neq n} \frac{R_{n,l(s,p)}^* \langle l|f_3|m\rangle \langle m|f_2|r\rangle \langle r|f_1|n\rangle}{(E_n^{(0)} - E_l^{(0)})(\omega_{m,n} - 2\omega_{inc})(\omega_{r,n} - \omega_{inc})} +$$

$$+ \sum_{m,r,l \neq n} \frac{R_{n,l,(s,p)}^* \langle l|f_2|m\rangle \langle m|f_3|r\rangle \langle r|f_1|n\rangle}{(E_n^{(0)} - E_l^{(0)})(\omega_{m,n} + \omega_{scat} - \omega_{inc})(\omega_{r,n} - \omega_{inc})} +$$

$$+ \sum_{m,r,l \neq n} \frac{R_{n,l,(s,p)}^* \langle l|f_2|m\rangle \langle m|f_1|r\rangle \langle r|f_3|n\rangle}{(E_n^{(0)} - E_l^{(0)})(\omega_{m,n} + \omega_{scat} - \omega_{inc})(\omega_{r,n} + \omega_{scat})} \tag{13}$$

where under $f_1, f_2, f_3$ we mean the dipole and quadrupole moments.

Using the following property of the scattering tensor

$$C_{V_{(s,p)}}[f_i, f_j, (a_1 f_k + a_2 f_m)] = a_1 C_{V_{(s,p)}}[f_i, f_j, f_k] + a_2 C_{V_{(s,p)}}[f_i, f_j, f_m]$$

$$C_{V_{(s,p)}}[f_i, (a_1 f_j + a_2 f_k), f_m] = a_1 C_{V_{(s,p)}}[f_i, f_j, f_m] + a_2 C_{V_{(s,p)}}[f_i, f_k, f_m]$$

$$C_{V_{(s,p)}}[(a_1 f_i + a_2 f_j), f_k, f_m] = a_1 C_{V_{(s,p)}}[f_i, f_k, f_m] + a_2 C_{V_{(s,p)}}[f_j, f_k, f_m] \tag{14}$$

the SEHR$_{(s,p)}$ cross-section can be written in the form

$$d\sigma_{SEHRS(s,p)}\begin{Bmatrix} St \\ AnSt \end{Bmatrix} = \frac{\omega_{inc}\omega_{scat}^3}{64\pi^2\hbar^4\varepsilon_0^2 c^4}\left|\overline{E}_{inc}\right|_{vol}^2 \frac{\left|\overline{E}_{inc}\right|_{surf}^2\left|\overline{E}_{inc}\right|_{surf}^2\left|\overline{E}_{scat}\right|_{surf}^2}{\left|\overline{E}_{inc}\right|_{vol}^2\left|\overline{E}_{inc}\right|_{vol}^2\left|\overline{E}_{scat}\right|_{vol}^2} \times$$

$$\times \left(\frac{\frac{V_{(s,p)}+1}{2}}{\frac{V_{(s,p)}}{2}}\right)\left|\sum_{f_1,f_2,f_3} S_{f_1-f_2-f_3}\right|^2 dO,$$

(15)

where

$$S_{d-d-d} = \sum_{i,j,k} C_{V_{(s,p)}}\left[d_i,d_j,d_k\right](e_{scat,i}^* e_{inc,j} e_{inc,k})_{surf} \quad (16)$$

$$S_{d-d-Q} = \sum_{i,j,\chi,\eta} C_{V_{(s,p)}}\left[d_i,d_j,Q_{\chi\eta}\right]\left(e_{scat,i}^* e_{inc,j}\frac{1}{2|\overline{E}_{inc}|}\frac{\partial E_\chi^{inc}}{\partial x_\eta}\right)_{surf} \quad (17)$$

$$S_{d-Q-d} = \sum_{i,\gamma,\delta,k} C_{V_{(s,p)}}\left[d_i,Q_{\gamma\delta},d_k\right]\left(e_{scat,i}^* \frac{1}{2|\overline{E}_{inc}|}\frac{\partial E_\gamma^{inc}}{\partial x_\delta} e_{inc,k}\right)_{surf} \quad (18)$$

$$S_{Q-d-d} = \sum_{\alpha,\beta,j,k} C_{V_{(s,p)}}[Q_{\alpha\beta},d_j,d_k]\left(\frac{1}{2|\overline{E}_{scat}|}\frac{\partial E_\alpha^{scat^*}}{\partial x_\beta} e_{inc,j} e_{inc,k}\right)_{surf} \quad (19)$$

$$S_{d-Q-Q} = \sum_{i,\gamma,\delta,\chi,\eta} C_{V_{(s,p)}}\left[d_i,Q_{\gamma\delta},Q_{\chi\eta}\right]\left(e_{scat,i}^* \frac{1}{2|\overline{E}_{inc}|}\frac{\partial E_\gamma^{inc}}{\partial x_\delta}\frac{1}{2|\overline{E}_{inc}|}\frac{\partial E_\chi^{inc}}{\partial x_\eta}\right)_{surf} \quad (20)$$

$$S_{Q-d-Q} = \sum_{j,\alpha,\beta,\chi,\eta} C_{V_{(s,p)}}\left[Q_{\alpha\beta},d_j,Q_{\chi\eta}\right]\left(\frac{1}{2|\overline{E}_{inc}|}\frac{\partial E_\alpha^{scat^*}}{\partial x_\beta} e_{inc,j}\frac{1}{2|\overline{E}_{inc}|}\frac{\partial E_\chi^{inc}}{\partial x_\eta}\right)_{surf} \quad (21)$$

$$S_{Q-Q-d} = \sum_{\alpha,\beta,\gamma,\delta,k} C_{V_{(s,p)}}[Q_{\alpha\beta},Q_{\gamma\delta},d_k]\left(\frac{1}{2|\overline{E}_{scat}|}\frac{\partial E_\alpha^{scat^*}}{\partial x_\beta}\frac{1}{2|\overline{E}_{inc}|}\frac{\partial E_\gamma^\delta}{\partial x_\delta} e_{inc,k}\right)_{surf} \quad (22)$$

$$S_{Q-Q-Q} = \sum_{\alpha,\beta,\gamma,\delta,\chi,\eta} C_{V_{(s,p)}} [Q_{\alpha\beta}, Q_{\gamma\delta}, Q_{\chi\eta}] \left( \frac{1}{2|\overline{E}_{scat}|} \frac{\partial E_{\alpha}^{scat^*}}{\partial x_{\beta}} \frac{1}{2|\overline{E}_{inc}|} \frac{\partial E_{\gamma}^{inc}}{\partial x_{\delta}} \frac{1}{2|\overline{E}_{inc}|} \frac{\partial E_{\chi}^{inc}}{\partial x_{\eta}} \right)_{surf}$$
(23)

are the sums of scattering contributions which occur via various dipole and quadrupole moments. One should note, that in each formula in (16-23) we have in fact the sum of various scattering contributions, which occur via various moments. Further it is convinient to transfer to combinations of the moments, which transform after irreducible representations of the molecule symmetry group. As it follows from the tables of irreducible representations[1], all the dipole and quadrupole moments $Q_{\alpha\beta}$ with $\alpha \neq \beta$ transform after irreducible representations in a major part of the point groups, while the $Q_{\alpha\alpha}$ moments can transform after reducible representations in these groups. Now we shall consider only this situation in order to simplify our investigations. Further it is convinient to transfer to linear combinations of the $Q_{\alpha\alpha}$ moments, which transform after irreducible representations. The general form of this transformation has the form

$$Q_1 = b_{11}Q_{xx} + b_{12}Q_{yy} + b_{13}Q_{zz},$$

$$Q_2 = b_{21}Q_{xx} + b_{22}Q_{yy} + b_{23}Q_{zz},$$

$$Q_3 = b_{31}Q_{xx} + b_{32}Q_{yy} + b_{33}Q_{zz}.$$
(24)

Then the $Q_{\alpha\alpha}$ moments are expressed via $Q_1, Q_2, Q_3$

$$Q_{xx} = a_{11}Q_1 + a_{12}Q_2 + a_{13}Q_3,$$

$$Q_{yy} = a_{21}Q_1 + a_{22}Q_2 + a_{23}Q_3,$$

$$Q_{zz} = a_{31}Q_1 + a_{32}Q_2 + a_{33}Q_3.$$
(25)

There will be combinations with a constant, purely positive sign, transforming after a unitary irreducible representation, which we shall call as the main quadrupole moments and combinations with a changeable sign, which are the minor ones. After substitution of (25) into (15), the SEHR$_{(s,p)}$ cross-section transfers to the form

$$d\sigma_{SEHRS(s,p)}\begin{pmatrix}St\\AnSt\end{pmatrix} = \frac{\omega_{inc}\omega_{scat}^3}{64\pi^2\hbar^4\varepsilon_0^2 c^4}\left|\overline{E}_{inc}\right|_{vol}^2 \frac{\left|\overline{E}_{inc}\right|_{surf}^2\left|\overline{E}_{inc}\right|_{surf}^2\left|\overline{E}_{scat}\right|_{surf}^2}{\left|\overline{E}_{inc}\right|_{vol}^2\left|\overline{E}_{inc}\right|_{vol}^2\left|\overline{E}_{scat}\right|_{vol}^2}\times$$

$$\times\left(\frac{\frac{V_{(s,p)}+1}{2}}{\frac{V_{(s,p)}}{2}}\right)\left|\sum_{f_1,f_2,f_3} T_{f_1-f_2-f_3}\right|^2 dO,$$

(26)

where

$$T_{d-d-d} = S_{d-d-d},\qquad(27)$$

$$T_{d-d-Q} = \sum_{\substack{i,j,\chi,\eta\\\chi\neq\eta}} C_{V_{(s,p)}}[d_i,d_j,Q_{\chi\eta}]\left(e_{scat,i}^* e_{inc,j}\frac{1}{2|\overline{E}_{inc}|}\frac{\partial E_\chi^{inc}}{\partial x_\eta}\right)_{surf} +$$

$$+ \sum_{i,j,\chi,k} a_{\chi,k} C_{V_{(s,p)}}[d_i,d_j,Q_k]\left(e_{scat,i}^* e_{inc,j}\frac{1}{2|\overline{E}_{inc}|}\frac{\partial E_\chi^{inc}}{\partial x_\chi}\right)_{surf}\qquad(28)$$

$$T_{d-Q-d} = \sum_{\substack{i,\gamma,\delta,k\\\gamma\neq\delta}} C_{V_{(s,p)}}[d_i,Q_{\gamma\delta},d_k]\left(e_{scat,i}^* \frac{1}{2|\overline{E}_{inc}|}\frac{\partial E_\gamma^{inc}}{\partial x_\delta} e_{inc,k}\right)_{surf} +$$

$$+ \sum_{i,\gamma,j,k} a_{\gamma,j} C_{V_{(s,p)}}[d_i,Q_j,d_k]\left(e_{scat,i}^* \frac{1}{2|\overline{E}_{inc}|}\frac{\partial E_\gamma^{inc}}{\partial x_\gamma} e_{inc,k}\right)_{surf}\qquad(29)$$

$$T_{Q-d-d} = \sum_{\substack{\alpha,\beta,j,k\\\alpha\neq\beta}} C_{V_{(s,p)}}[Q_{\alpha\beta},d_j,d_k]\left(\frac{1}{2|\overline{E}_{scat}|}\frac{\partial E_\alpha^{scat^*}}{\partial x_\beta} e_{inc,j} e_{inc,k}\right) +$$

$$+ \sum_{\alpha,i,j,k} a_{\alpha,i} C_{V_{(s,p)}}[Q_i,d_j,d_k]\left(\frac{1}{2|\overline{E}_{scat}|}\frac{\partial E_\alpha^{scat^*}}{\partial x_\alpha} e_{inc,j} e_{inc,k}\right)\qquad(30)$$

$$T_{d-Q-Q} = \sum_{\substack{i,\gamma,\delta,\chi,\eta\\\gamma\neq\delta\\\chi\neq\eta}} C_{V_{(s,p)}}[d_i,Q_{\gamma\delta},Q_{\chi\eta}]\left(e_{scat,i}^* \frac{1}{2|\overline{E}_{inc}|}\frac{\partial E_\gamma^{inc}}{\partial x_\delta}\frac{1}{2|\overline{E}_{inc}|}\frac{\partial E_\chi^{inc}}{\partial x_\eta}\right)_{surf} +$$

$$+ \sum_{\substack{i,\gamma,\delta,\chi,k \\ \gamma \neq \delta}} a_{\gamma,k} C_{V_{(s,p)}} \left[ d_i, Q_{\gamma\delta}, Q_k \right] \left( e_{scat,i}^* \frac{1}{2|\overline{E}_{inc}|} \frac{\partial E_\gamma^{inc}}{\partial x_\delta} \frac{1}{2|\overline{E}_{inc}|} \frac{\partial E_\chi^{inc}}{\partial x_\chi} \right)_{surf} +$$

$$+ \sum_{i,\gamma,j,\chi,\eta} a_{\gamma,j} C_{V_{(s,p)}} \left[ d_i, Q_j, Q_{\chi\eta} \right] \left( e_{scat,i}^* \frac{1}{2|\overline{E}_{inc}|} \frac{\partial E_\gamma^{inc}}{\partial x_\gamma} \frac{1}{2|\overline{E}_{inc}|} \frac{\partial E_\chi^{inc}}{\partial x_\eta} \right)_{surf} +$$

$$+ \sum_{i,\gamma,j,\chi,k} a_{\gamma,j} a_{\chi,k} C_{V_{(s,p)}} \left[ d_i, Q_j, Q_k \right] \left( e_{scat,i}^* \frac{1}{2|\overline{E}_{inc}|} \frac{\partial E_\gamma^{inc}}{\partial x_\gamma} \frac{1}{2|\overline{E}_{inc}|} \frac{\partial E_\chi^{inc}}{\partial x_\chi} \right)_{surf} \quad (31)$$

$$T_{Q-d-Q} = \sum_{\substack{\alpha,\beta,j,\gamma,\delta \\ \alpha \neq \beta \\ \gamma \neq \delta}} C_{V_{(s,p)}} \left[ Q_{\alpha\beta}, d_j, Q_{\gamma\delta} \right] \left( \frac{1}{2|\overline{E}_{inc}|} \frac{\partial E_\alpha^{scat^*}}{\partial x_\beta} e_{inc,j} \frac{1}{2|\overline{E}_{inc}|} \frac{\partial E_\gamma^{inc}}{\partial x_\delta} \right)_{surf} +$$

$$+ \sum_{\substack{\alpha,\beta,j,\gamma,k \\ \alpha \neq \beta}} a_{\gamma,k} C_{V_{(s,p)}} \left[ Q_{\alpha\beta}, d_j, Q_k \right] \left( \frac{1}{2|\overline{E}_{inc}|} \frac{\partial E_\alpha^{scat^*}}{\partial x_\beta} e_{inc,j} \frac{1}{2|\overline{E}_{inc}|} \frac{\partial E_\gamma^{inc}}{\partial x_\gamma} \right)_{surf} +$$

$$+ \sum_{\substack{\alpha,i,j,\chi,\eta \\ \chi \neq \eta}} a_{\alpha,i} C_{V_{(s,p)}} \left[ Q_i, d_j, Q_{\chi\eta} \right] \left( \frac{1}{2|\overline{E}_{inc}|} \frac{\partial E_\alpha^{scat^*}}{\partial x_\alpha} e_{inc,j} \frac{1}{2|\overline{E}_{inc}|} \frac{\partial E_\chi^{inc}}{\partial x_\eta} \right)_{surf} +$$

$$+ \sum_{\alpha,i,j,\chi,k} a_{\alpha,i} a_{\chi,k} C_{V_{(s,p)}} \left[ Q_i, d_j, Q_k \right] \left( \frac{1}{2|\overline{E}_{inc}|} \frac{\partial E_\alpha^{scat^*}}{\partial x_\alpha} e_{inc,j} \frac{1}{2|\overline{E}_{inc}|} \frac{\partial E_\chi^{inc}}{\partial x_\chi} \right)_{surf} \quad (32)$$

$$T_{Q-Q-d} = \sum_{\substack{\alpha,\beta,\gamma,\delta,k \\ \alpha \neq \beta \\ \gamma \neq \delta}} C_{V_{(s,p)}} \left[ Q_{\alpha\beta}, Q_{\gamma\delta}, d_k \right] \left( \frac{1}{2|\overline{E}_{inc}|} \frac{\partial E_\alpha^{scat^*}}{\partial x_\beta} \frac{1}{2|\overline{E}_{inc}|} \frac{\partial E_\gamma^{inc}}{\partial x_\delta} e_{inc,k} \right)_{surf} +$$

$$+ \sum_{\substack{\alpha,\beta,\gamma,j,k \\ \alpha \neq \beta}} a_{\gamma,j} C_{V_{(s,p)}} \left[ Q_{\alpha\beta}, Q_j, d_k \right] \left( \frac{1}{2|\overline{E}_{inc}|} \frac{\partial E_\alpha^{scat^*}}{\partial x_\beta} \frac{1}{2|\overline{E}_{inc}|} \frac{\partial E_\gamma^{inc}}{\partial x_\gamma} e_{inc,k} \right)_{surf} +$$

$$+ \sum_{\substack{\alpha,i,\gamma,\delta,k \\ \gamma \neq \delta}} a_{\alpha,i} C_{V_{(s,p)}} \left[ Q_i, Q_{\gamma\delta}, d_k \right] \left( \frac{1}{2|\overline{E}_{inc}|} \frac{\partial E_\alpha^{scat^*}}{\partial x_\alpha} \frac{1}{2|\overline{E}_{inc}|} \frac{\partial E_\gamma^{inc}}{\partial x_\delta} e_{inc,k} \right)_{surf} +$$

$$+ \sum_{\alpha,i,\gamma,j,k} a_{\alpha,i} a_{\gamma,j} C_{V_{(s,p)}} \left[ Q_i, Q_j, d_k \left( \frac{1}{2|\overline{E}_{inc}|} \frac{\partial E_\alpha^{scat^*}}{\partial x_\alpha} \frac{1}{2|\overline{E}_{inc}|} \frac{\partial E_\gamma^{inc}}{\partial x_\gamma} e_{inc,k} \right)_{surf} \right] \qquad (33)$$

$$T_{Q-Q-Q} = \sum_{\substack{\alpha,\beta,\gamma,\delta,\chi,\eta \\ \alpha \neq \beta \\ \gamma \neq \delta \\ \chi \neq \eta}} C_{V_{(s,p)}} \left[ Q_{\alpha\beta}, Q_{\gamma\delta}, Q_{\chi\eta} \left( \frac{1}{2|\overline{E}_{scat}|} \frac{\partial E_\alpha^{scat^*}}{\partial x_\beta} \frac{1}{2|\overline{E}_{inc}|} \frac{\partial E_\gamma^{inc}}{\partial x_\delta} \frac{1}{2|\overline{E}_{inc}|} \frac{\partial E_\chi^{inc}}{\partial x_\eta} \right)_{surf} \right] +$$

$$+ \sum_{\substack{\alpha,\beta,\gamma,\delta,\chi,k \\ \alpha \neq \beta \\ \gamma \neq \delta}} a_{\chi,k} C_{V_{(s,p)}} \left[ Q_{\alpha\beta}, Q_{\gamma\delta}, Q_k \left( \frac{1}{2|\overline{E}_{scat}|} \frac{\partial E_\alpha^{scat^*}}{\partial x_\beta} \frac{1}{2|\overline{E}_{inc}|} \frac{\partial E_\gamma^{inc}}{\partial x_\delta} \frac{1}{2|\overline{E}_{inc}|} \frac{\partial E_\chi^{inc}}{\partial x_\chi} \right)_{surf} \right] +$$

$$+ \sum_{\substack{\alpha,\beta,\gamma,j,\chi,\eta \\ \alpha \neq \beta \\ \chi \neq \eta}} a_{\gamma,j} C_{V_{(s,p)}} \left[ Q_{\alpha\beta}, Q_j, Q_{\chi\eta} \left( \frac{1}{2|\overline{E}_{scat}|} \frac{\partial E_\alpha^{scat^*}}{\partial x_\beta} \frac{1}{2|\overline{E}_{inc}|} \frac{\partial E_\gamma^{inc}}{\partial x_\gamma} \frac{1}{2|\overline{E}_{inc}|} \frac{\partial E_\chi^{inc}}{\partial x_\eta} \right)_{surf} \right] +$$

$$+ \sum_{\substack{\alpha,i,\gamma,\delta,\chi,\eta \\ \gamma \neq \delta \\ \chi \neq \eta}} a_{\alpha,i} C_{V_{(s,p)}} \left[ Q_i, Q_{\gamma\delta}, Q_{\chi\eta} \left( \frac{1}{2|\overline{E}_{scat}|} \frac{\partial E_\alpha^{scat^*}}{\partial x_\alpha} \frac{1}{2|\overline{E}_{inc}|} \frac{\partial E_\gamma^{inc}}{\partial x_\delta} \frac{1}{2|\overline{E}_{inc}|} \frac{\partial E_\chi^{inc}}{\partial x_\eta} \right)_{surf} \right] +$$

$$+ \sum_{\substack{\alpha,\beta,\gamma,j,\chi,k \\ \alpha \neq \beta}} a_{\gamma,j} a_{\chi,k} C_{V_{(s,p)}} \left[ Q_{\alpha\beta}, Q_j, Q_k \left( \frac{1}{2|\overline{E}_{scat}|} \frac{\partial E_\alpha^{scat^*}}{\partial x_\beta} \frac{1}{2|\overline{E}_{inc}|} \frac{\partial E_\gamma^{inc}}{\partial x_\gamma} \frac{1}{2|\overline{E}_{inc}|} \frac{\partial E_\chi^{inc}}{\partial x_\chi} \right)_{surf} \right] +$$

$$+ \sum_{\substack{\alpha,i,\gamma,\delta,\chi,k \\ \gamma \neq \delta}} a_{\alpha,i} a_{\chi,k} C_{V_{(s,p)}} \left[ Q_i, Q_{\gamma\delta}, Q_k \left( \frac{1}{2|\overline{E}_{scat}|} \frac{\partial E_\alpha^{scat^*}}{\partial x_\alpha} \frac{1}{2|\overline{E}_{inc}|} \frac{\partial E_\gamma^{inc}}{\partial x_\delta} \frac{1}{2|\overline{E}_{inc}|} \frac{\partial E_\chi^{inc}}{\partial x_\chi} \right)_{surf} \right] +$$

$$+ \sum_{\substack{\alpha,i,\gamma,j,\chi,\eta \\ \chi \neq \eta}} a_{\alpha,i} a_{\gamma,j} C_{V_{(s,p)}} \left[ Q_i, Q_j, Q_{\chi\eta} \left( \frac{1}{2|\overline{E}_{scat}|} \frac{\partial E_\alpha^{scat^*}}{\partial x_\alpha} \frac{1}{2|\overline{E}_{inc}|} \frac{\partial E_\gamma^{inc}}{\partial x_\gamma} \frac{1}{2|\overline{E}_{inc}|} \frac{\partial E_\chi^{inc}}{\partial x_\eta} \right)_{surf} \right] +$$

$$+ \sum_{\alpha,i,\gamma,j,\chi,k} a_{\alpha,i} a_{\gamma,j} a_{\chi,k} C_{V_{(s,p)}} [Q_i, Q_j, Q_k] \left( \frac{1}{2|\overline{E}_{scat}|} \frac{\partial E_\alpha^{scat*}}{\partial x_\alpha} \frac{1}{2|\overline{E}_{inc}|} \frac{\partial E_\gamma^{inc}}{\partial x_\gamma} \frac{1}{2|\overline{E}_{inc}|} \frac{\partial E_\chi^{inc}}{\partial x_\chi} \right)_{surf}$$

(34)

Here $T$ are the sums of transformed contributions, which depend on the dipole and quadrupole moments of some definite type.

## 2. Estimation of the enhancement of the dipole and quadrupole interactions in SEHRS

As it was demonstrated in[1-7] the enhancement of the dipole and quadrupole interactions arises due to huge increase of the $E_z$ component of the electric field, which is perpendicular to the surface and derivatives of this field $\partial E_\alpha / \partial x_\alpha$ near sharp features of the surface. Another essential reason of the enhancement, of the quadrupole interaction is a quantum-mechanical feature of matrix elements of the quadrupole moments $Q_{\alpha\alpha}$ with a constant sign. The matrix elements of these moments are significantly larger, than the matrix elements of the moments $Q_{\alpha\beta}$. The Hamiltonian of interaction of light with molecule consists of the dipole and quadrupole parts (3-5):

$$\hat{H}_{e-r} = \hat{H}_d + \hat{H}_Q \tag{35}$$

A crude estimation of the enhancement of separate terms of the quadrupole interaction relative to those of the dipole interaction in a free space can be obtained from (3,4) using (5)

$$\frac{\langle m|Q_{\alpha\beta}|n\rangle}{\langle m|d_\alpha|n\rangle} \frac{1}{2(E_\alpha)_{vol}} \left(\frac{\partial E_\alpha}{\partial x_\beta}\right)_{surf} = B_{\alpha\beta} a \frac{1}{2(E_\alpha)_{vol}} \left(\frac{\partial E_\alpha}{\partial x_\beta}\right)_{surf} \tag{36}$$

The first factor in the left side of (36) is the relation of some mean matrix elements of the quadrupole and dipole transitions, which is equal to $B_{\alpha\beta} a$. Here $a$ is a molecule size, $B_{\alpha\beta}$ are some numerical coefficients. The $B_{\alpha\beta}$ values essentially differ for $\alpha \neq \beta$ and for $\alpha = \beta$, since $d_\alpha$ and $Q_{\alpha\beta}$ are the values with a changeable sign, while $Q_{\alpha\alpha}$ are the ones with a constant sign, that strongly increases the $B_{\alpha\alpha}$ values. The first factor in the left side of (36) reflects the large increase of the matrix elements of the quadrupole moments and transitions due to $Q_{\alpha\alpha}$, with respect to those, caused by the dipole moments $d_\alpha$. For estimation of the $B_{\alpha\alpha}a$ values it is necessary to make the following simplifications. Because the inner electron shell of molecules for excited states remains almost intact, then for estimations we usually take the value $\overline{\langle n|Q_{,\alpha\alpha}|n\rangle}$ instead of $\overline{\langle n|Q_{,\alpha\alpha}|l\rangle}$ in (36) and the value $\sqrt{e^2\hbar/2m\omega_{nl}} \times \sqrt{\overline{f}_{nl}}$ for $\overline{\langle n|d_{,\alpha}|l\rangle}$, which is expressed in terms of some mean value of the oscillator strength $\overline{f}_{nl} = 0.1$ while $\omega_{nl}$ corresponds to the edge of absorption (approximately $\lambda = 2500 A$ for the pyridine molecule for example). The first approximation is forced because of the absence of data on quadrupole transitions. Since the configuration of the electron shell is close to the configuration of nuclei the value $\overline{\langle n|Q_{\alpha\alpha}|n\rangle}$ was calculated as a $Q_{n,\alpha\alpha}$ component of the quadrupole moments of nuclei. Then estimation of $B_{\alpha\alpha}$ for the pyridine, benzene or pyrazine molecules give the value $B_{\alpha\alpha} \sim 2 \times 10^2$, that strongly differs from the value $B_{\alpha\alpha} \sim 1$ ($B_{\alpha\alpha}a \sim a$). The last one is usually used in literature as the relation of the quadrupole and dipole operators. Taking into account the dependence of the radial component of the electric field in the vicinity of the top of a model feature of a real roughness of a cone type (Fig. 1)

$$E_r \sim \left|\overline{E}_{inc}\right|_{vol} C_0 \left(\frac{l_1}{r}\right)^\beta, \tag{37}$$

the estimation of the relation of the quadrupole and dipole interactions results in the following expressions for the enhancement coefficients of the quadrupole and dipole terms of Hamiltonian

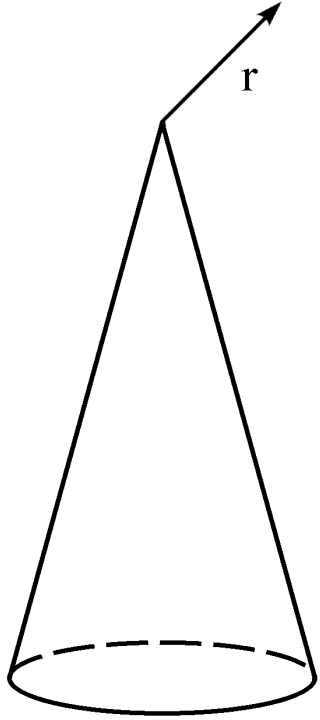

Figure 1. The roughness of the cone type

$G_{H_Q}$ and $G_{H_d}$

$$G_{H_Q} \sim C_0 \beta \left(\frac{B_{\alpha\alpha}}{2}\right)\left(\frac{l_1}{r}\right)^\beta \left(\frac{a}{r}\right) \qquad (38)$$

$$G_{H_d} \sim C_0 \left(\frac{l_1}{r}\right)^\beta \qquad (39)$$

Here $C_0$ is a numerical coefficient, $0 < \beta < 1$ and depends on the cone angle at the top of the cone, $l_1$ is a characteristic size of the cone, $r$ is a radius vector. It is seen that for reasonable values $C_0 \sim 1$, $l_1 \sim 10$ nm, $r \sim 1$ nm, $\beta \sim 1$ and the molecules like pyridine, benzene or pyrazine with $B_{\alpha\alpha} \sim 2 \times 10^2$ the enhancement of the dipole interaction is $\sim 10$, while the enhancement of the quadrupole interaction is $\sim 10^2$.

The enhancement in SEHRS can be estimated using usual ideas, that this process is the optical process of the third order. Therefore the enhancement coefficient in SEHRS for pure quadrupole interaction can be estimated as

$$G_Q \sim C_0^6 \beta^6 \left(\frac{B_{\alpha\alpha}}{2}\right)^6 \left(\frac{l_1}{r}\right)^{6\beta} \left(\frac{a}{r}\right)^6 \qquad (40)$$

while the estimation of the enhancement coefficient due to the pure dipole interaction is

$$G_d \sim C_0^6 \left(\frac{l_1}{r}\right)^{6\beta} \qquad (41)$$

For the values of the parameters pointed out above, the enhancement due to the purely dipole interaction is of the order of $10^6$, while the enhancement due to the quadrupole interaction $\sim 10^{12}$. One should note that the enhancement of the dipole and quadrupole interactions $G_{H_d}$ and $G_{H_Q}$ and especially the enhancement in SEHRS $G_Q$ can be very large. For example for some limited situations with the values of the parameters $C_0 \sim 1$, $B_{\alpha\alpha} \sim 2 \times 10^2$, $r \sim 0.1 nm$, $\beta \sim 1$, $l_1 \sim 100 nm$ corresponding to the placement of the molecule on the top of the cone (tip or spike) the enhancement $G_Q$ in SEHRS may achieve $10^{30}$. As it was pointed out in[1] for example, the real situation is that most enhancement arises in the vicinity of some points associated with prominent places with a very large curvature. The mean enhancement is formed from the whole layer of adsorbed molecules and is significantly smaller than the maximum enhancement near these places due to averaging.

In accordance with the above consideration all the moments can be separated in two groups, those which are responsible for the large enhancement and those, which are nonessential for the enhancement. This classification depends on the coverage of substrate. For monolayer coverage only the $d_z$ moment, which is perpendicular to the surface and $Q_{xx}, Q_{yy}, Q_{zz}$ quadrupole moments are responsible for the strong enhancement. Therefore we shall name them as main moments. All others we shall name as the minor ones. For multilayer coverage, all the $d$ moments can be essential for the enhancement[1], since the molecules can be oriented in arbitrary manner on the substrate due to superposition of molecules in the first and possible arbitrary orientation in the second and upper layers. Therefore all the $d$ moments can refer to the main moments in this case.

### 3. Selection rules for the contributions

In accordance with (26) the cross-section is expressed via the sum of the contributions $T_{f_1-f_2-f_3}$, which depend on three dipole and quadrupole moments transforming after irreducible representations of the symmetry group. In accordance with the expressions (27-34) these contributions are not equal to zero when

$$C_{V_{(s,p)}}[f_1, f_2, f_3] \neq 0 \tag{42}$$

Let us consider one, the first line in the first sum in (13). The condition

$$R_{n,l,(s,p)} \langle n|f_1|m \rangle \langle m|f_2|r \rangle \langle r|f_1|l \rangle \neq 0 \tag{43}$$

is valid, when the following conditions

$$R_{n,l,(s,p)} \neq 0 \tag{44}$$

$$\langle r|f_1|l \rangle \neq 0 \tag{45}$$

$$\langle m|f_2|r \rangle \neq 0 \tag{46}$$

$$\langle n|f_3|m \rangle \neq 0 \tag{47}$$

are fulfilled. Let us designate the irreducible representations which determine transformational properties of the $(s, p)$ vibration and of the dipole and quadrupole moments $f$ by symbols $\Gamma$. The expression (44) is valid when[1]

$$\Gamma_{(s,p)} \in \Gamma_l \Gamma_n, \tag{48}$$

while other expressions can be used for determination of the irreducible representations for the wavefunctions in their right side,

$$\Gamma_l \in \Gamma_r \Gamma_{f_1} \ , \quad \Gamma_r \in \Gamma_m \Gamma_{f_2} \quad \text{and} \quad \Gamma_m \in \Gamma_n \Gamma_{f_3} \tag{49}$$

After the consecutive substitution of the expressions from (49) in (48) one can obtain the following condition, when (43) is valid

$$\Gamma_{(s,p)} \in \Gamma_{f_1} \Gamma_{f_2} \Gamma_{f_3} \tag{50}$$

Analysis of another lines in (13) results in the same expression (50). Thus the expression (50) presents selection rules for the $T_{f_1-f_2-f_3}$ contributions.

Here we have obtained the selection rules, which are valid in the symmetry groups, pointed out above. The condition was that the dipole and quadrupole moments $Q_{\alpha\beta}$, $(\alpha \neq \beta)$ transform after irreducible representations. For the case of the groups where the dipole and quadrupole moments $Q_{\alpha\beta}$ transform after reducible representations one can transfer to the combinations of these moments, transforming after irreducible representations. Then after transformation of the cross-section (15), similar to those, made with the moments $Q_{\alpha\alpha}$, one can obtain the expression for the SEHRS cross-section that coincides formally with (26) with the $T_{f_1-f_2-f_3}$ contributions which slightly differ from those obtained for the previous case. All the $f_1, f_2$ and $f_3$ moments are now combinations of the dipole and quadrupole moments, which transform after irreducible representations of the corresponding symmetry groups and the selection rules (50) are valid now for all point groups. The definition of the main and minor quadrupole moments remains the same and is determined by the constancy or changeability of the sign, while the definition of the main and minor dipole moments depends on the coverage of substrate and is defined in the following manner. The $d_z$ moment is the main one, while the linear combinations of $d_x$ and $d_y$ moments are the minor ones for monolayer coverage. For multilayer coverage the $d_z$ and the linear combinations of $d_x$ and $d_y$ moments can be the main moments, because of possible arbitrary orientation of molecules near substrate.

## 4. Classification of the contributions in accordance with enhancement degree

In accordance with our previous consideration the most enhancement for the strongly rough surface is caused by the quadrupole interaction with $Q_{main}$ moments and by the dipole interaction with the dipole moment $d_z$, which is perpendicular to the surface. Than the contributions $T_{f_1-f_2-f_3}$ which we shall designate further simply as $(f_1 - f_2 - f_3)$ can be classified qualitatively after the enhancement degree in the following manner:

1. $(Q_{main} - Q_{main} - Q_{main})$ - the most enhanced scattering type.

2. $(Q_{main} - Q_{main} - d_z)$ - scattering type, which can be strongly enhanced too, but in a lesser degree than the previous one.

3. $(Q_{main} - d_z - d_z)$ - scattering type, which can be strongly enhanced too, but lesser, than the two previous ones and

4. $(d_z - d_z - d_z)$ - scattering type, which can be strongly enhanced too, but lesser than the three previous ones.

Here and further we mean under $(f_1 - f_2 - f_3)$ all contributions with permutations of the $f$ moments. Considering molecules with $C_{nh}, D$ and higher symmetry one can note, that the first and the third enhancement types contribute to the bands, caused by vibrations transforming after the unit irreducible representation or by the totally symmetric vibrations, while the second and the forth types contribute to the bands, caused by vibrations transforming such as the $d_z$ moment. Thus the most enhanced bands in molecules with $C_{nh}, D$ and higher symmetry groups are caused by the above types of vibrations. The first ones are forbidden in usual HRS. Thus these bands must be an essential feature of the SEHR spectra of symmetrical molecules with the above symmetry. The above consideration deals with the molecules with some definite orientation of the molecule with respect to the surface, when the chosen orientation of the $d_z$ moment of the molecule coincides with the direction of the $E_z$ component of the electric field. However sometimes molecules can be arbitrary oriented at the surface, because of their superposition. Then all the $d$ moments can contribute to the scattering. Thus the contributions of $(Q_{main} - Q_{main} - d_i)$ and $(Q_{main} - d_i - d_k)$ scattering types can manifest in the SEHR spectra in the symmetry groups, where $d_x$ and $d_y$ moments transform after irreducible representations. Apparently the contributions of the $(Q_{main} - Q_{main} - f_i)$ and $(Q_{main} - f_i - f_k)$ scattering types, where $f_i$ and $f_k$ are linear combinations of the $d_x$ and $d_y$ moments may manifest in the SEHRS spectra too. This notion refers to the molecules with the second type of symmetry groups, where $d_x$ and $d_y$ moments transform after reducible representations. The other contributions without $Q_{main}$, or those, which contain $Q_{min\,or}$ moments apparently will be small and may manifest in the SEHR spectra only in very strong incident fields.

One should specially note, that the experiments on SEHRS can be performed in solutions. It is another reason, when the molecules can be oriented arbitrary. As it was mentioned, all the $d$ moments can be essential for the scattering in this case and are the main ones. Our classification in this case has a form

1. $(Q_{main} - Q_{main} - Q_{main})$ - the most enhanced scattering type.

2. $(Q_{main} - Q_{main} - d_{main})$ - scattering type, which can be strongly enhanced too, but in a lesser degree than the previous one.

3. ( $Q_{main} - d_{main} - d_{main}$ )- scattering type, which can be strongly enhanced too, but lesser, than the two previous ones and

4. ( $d_{main} - d_{main} - d_{main}$ )- scattering type, which can be strongly enhanced too, but lesser than the three previous ones.

The scattering type of the first type and the third type with the same $d_{main}$ moments determine enhancement of the bands, caused by the totally symmetric vibrations as in the previous case. The contributions of the second type determine the enhancement of the bands, transforming after the same irreducible representation as the $d_{main}$ moment. Other contributions may define enhancement of the bands, which are forbidden in usual HRS. This reasoning is valid for the case of the groups, when $d_x$ and $d_y$ moments transform after irreducible representations. The contributions of the $(Q_{main} - Q_{main} - f_i)$ and $(Q_{main} - f_i - f_k)$ scattering types, where $f_i$ and $f_k$ are linear combinations of the $d_x$ and $d_y$ moments may manifest in the SEHRS spectra too.

## 5. Acknowledgement

In conclusion the author want to thanks professor V.P. Smirnov for valuable discussions